\documentclass[fp,twocolumn]{jpsj3}
\usepackage{txfonts}

\title{
Analytical Solutions for the Surface States of Bi$_{1-x}$Sb$_x$ ($0\le x \lesssim 0.1$)}

\author{Yuki Fuseya$^1$\thanks{fuseya@uec.ac.jp} and Hidetoshi Fukuyama$^2$}
\inst{$^1$Department of Engineering Science, University of Electro-Communications, Chofu, Tokyo 182-8585, Japan \\
$^2$Research Institute for Science and Technology, Tokyo University of Science, Kagurazaka, Shinjuku-ku, Tokyo 162-8601, Japan} 

\abst{Analytical solutions for the surface state (SS) of an extended Wolff Hamiltonian, which is a common Hamiltonian for strongly spin-orbit coupled systems, are obtained both for semi-infinite and finite-thickness boundary conditions. For the semi-infinite system, there are three types of SS solutions: (I-a) linearly crossing SSs in the direct bulk band gap, (I-b) SSs with linear dispersions entering the bulk conduction or valence bands away from the band edge, and (II) SSs with nearly flat dispersions entering the bulk state at the band edge. 
For the finite-thickness system, a gap opens in the SS of solution I-a. 
Numerical solutions for the SS are also obtained based on the tight-binding model of Liu and Allen [Phys. Rev. B, {\bf 52}, 1566 (1995)] for Bi$_{1-x}$Sb$_x$ ($0\le x \le 0.1$). A perfect correspondence between the analytic and numerical solutions is obtained around the $\bar{M}$ point including their thickness dependence. This is the first time that the character of the SS numerically obtained is identified with the help of analytical solutions.
The size of the gap for I-a SS can be larger than that of bulk band gap even for a ``thick" films ($\lesssim 200$ bilayers $\simeq 80$ nm) of pure bismuth. Consequently, in such a film of Bi$_{1-x}$Sb$_x$, there is no apparent change in the SSs through the band inversion at $x\simeq 0.04$, even though the nature of the SS is changed from solution I-a to I-b. Based on our theoretical results, the experimental results on the SS of Bi$_{1-x}$Sb$_x$ ($0\le x \lesssim 0.1$) are discussed.
}

\usepackage{amsmath,bm,mathrsfs,color}

\newcommand{\scr}[1]{\mathscr{#1}}

\newcommand{\bp}{\bm{p}}

\newcommand{\bW}{\bm{W}}

\newcommand{\kp}{\bm{k}\cdot \bm{p}}
\newcommand{\D}{\Delta}

\begin{document}
\maketitle
\section{Introduction}
Spin-orbit coupling (SOC) produces an unconventional surface state (SS) that is inseparable from the electronic states of the bulk solid the surface belongs to. This unconventional SS is very different from SSs that are separable from the bulk states, which can be due to dangling bonds and surface reconstruction.
From the viewpoint of the SOC-originated SS, bismuth is one of the most fascinating materials\cite{Hofmann2006,Hirahara2015,Jezequel1986,Hengsberger2000,Ast2001,Ast2003,Koroteev2004,Hirahara2006,Hirahara2007} because it has the largest SOC among the radioactively safe elements. Moreover, following the proposal of the existence of a three-dimensional topological insulator in Bi$_{1-x}$Sb$_x$\cite{Fu2007}, the SS of Bi$_{1-x}$Sb$_x$ (especially for $0\le x \lesssim 0.1$) has attracted considerable attention in solid-state physics.\cite{Hsieh2008,Teo2008,Hsieh2009,HJZhang2009,Hirahara2010,Nishide2010,HGuo2011,Nakamura2011,Ohtsubo2013,Benia2015,Aguilera2015,Ito2016,Ohtsubo2016}. While it has been widely accepted that pure Bi is topologically trivial, it has been proposed that Bi$_{1-x}$Sb$_x$ becomes topologically nontrivial\cite{Fu2007,Teo2008} due to band inversion at the $L$ point in a bulk Brillouin zone\cite{Jain1959,Lerner1968,Tichovolsky1969,Brandt1970,Buot1971,Brandt1972,Oelgart1976,Vecchi1976,Lenoir1996}. A drastic change in the SS is therefore expected due to this topological transition caused by the substitution of Sb into Bi.

\begin{figure}[ht]
\begin{center}
\includegraphics[width=7cm]{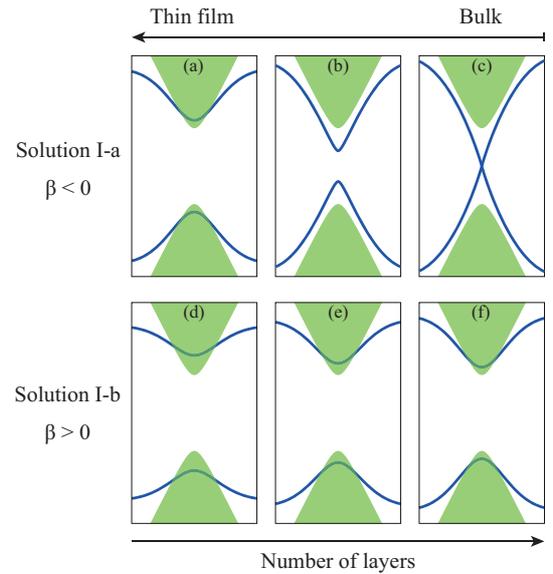}
\end{center}
\caption{\label{fig_illust_theory}Schematic summary of the present analytical solutions. The right panels (c) and (f) correspond to the semi-infinite system. The shaded region expresses the bulk band, the extremum of which corresponds to the $\bar{M}$ point in the surface Brillouin zone of Bi$_{1-x}$Sb$_x$.
}
\end{figure}

The (111) surfaces of Bi$_{1-x}$Sb$_x$ have been examined extensively. While it seems that the experimental results for SS are heading toward to be settled, the interpretation of the results is still controversial; the SS around the $\bar{M}$ point is of a particular interest. (The $\bar{M}$ point in a surface Brillouin zone corresponds to the $L$ point in a bulk Brillouin zone.) 
According to the previously obtained experimental results\cite{Hofmann2006,Hirahara2015,Jezequel1986,Hengsberger2000,Ast2001,Ast2003,Koroteev2004,Hirahara2006,Hirahara2007,Hsieh2008,Teo2008,Hsieh2009,HJZhang2009,Hirahara2010,Nishide2010,HGuo2011,Nakamura2011,Ohtsubo2013,Benia2015,Aguilera2015,Ito2016,Ohtsubo2016}, there are two SSs between the $\bar{\Gamma}$ and $\bar{M}$ points in a surface Brillouin zone. Near the $\bar{M}$ point, one SS enters into a bulk conduction band, while the other enters into the bulk valence band. 
The controversy centers around the fact that the SS does not exhibit a qualitative change with respect to the Sb substitution for $0\le x \lesssim 0.2$, even though the bulk conduction and valence bands at the $L$ point are inverted at $x\simeq 0.04$\cite{Jain1959,Lerner1968,Tichovolsky1969,Brandt1970,Buot1971,Brandt1972,Oelgart1976,Vecchi1976,Lenoir1996}, where the topology of the system is  changed\cite{Fu2007,Teo2008}. (It should be noted that in early studies on topological insulators, a third SS seemed to appear by the band inversion\cite{Hsieh2008,Hsieh2009,Nishide2010,HGuo2011,Nakamura2011}, however, recent experiments have attributed this third band to surface imperfections\cite{HJZhang2009,Benia2015}.) 

Because of these unexpected behaviors, it has recently been suggested that pure Bi should be topologically nontrivial\cite{Ohtsubo2013,Ito2016,Ohtsubo2016}. If pure Bi is topologically nontrivial, then Bi$_{1-x}$Sb$_x$ should become topologically trivial after the band inversion ($x\gtrsim 0.04$), namely, a qualitative change in the SS should occur through the band inversion after all, even if pure Bi is topologically nontrivial. However, the experimental observations that have been made do not indicate such a qualitative change\cite{Hirahara2010,Nishide2010,HGuo2011,Nakamura2011,Benia2015}.
Furthermore, the band structure of pure Bi was meticulously re-examined using first-principles calculations, and pure Bi was nevertheless determined to be topologically trivial\cite{Aguilera2015,Harima2017}.

The core of the mystery is at the properties of SS around the $\bar{M}$ point. Thus the problem should be solved by detailed investigation of the local region in the Brillouin zone. In this paper, we throughly investigate the SSs around the $\bar{M}$ point on the basis of $\kp$ theory, an approach that has never before been adopted for the current problem. We succeeded in obtaining analytical solutions for the effective Hamiltonian common to the strongly spin-orbit coupled system --- the Wolff Hamiltonian, the validity of which has been proved by the long history of studies on Bi and Bi-Sb alloys\cite{Cohen1960,Wolff1964,Fukuyama1970,Fuseya2015}. This approach is unbiased against the topological properties of the system\cite{Konig2008,BZhou2008,Linder2009,HZLu2010,CXLiu2010,WYShan2010,SQShen2012,SQShen_book}. 

The essence of our findings is as follows. For a semi-infinite system (i.e., the bulk limit), there are two solutions, I-a and I-b, which are responsible to the current the problem. In solution I-a, the two linear SS dispersions cross over one another in the direct bulk band gap; this is depicted in Fig. \ref{fig_illust_theory} (c). In solution I-b, on the other hand, one SS linearly enters into the bulk conduction band away from the band-edge, while the other enters into the bulk valence band as in Fig. \ref{fig_illust_theory} (f). This result has not been pointed out by the previous analytical approaches\cite{Konig2008,BZhou2008,Linder2009,HZLu2010,CXLiu2010,WYShan2010,SQShen2012,SQShen_book} and this study is the first to show it. Their existence condition  is given by the parameter $\beta=\alpha'_{zz}/\alpha_{zz}$, where $\alpha_{zz}$ is an inverse mass tensor that originates from the matrix elements between the bulk conduction and valence bands perpendicular to the surface, and $\alpha'_{zz}$ is that from the other bands. Solution I-a is obtained when $\beta <0$, while I-b is obtained when $\beta >0$. For a system of finite-thickness, however, a gap opens in solution I-a, as can be seen in Fig. \ref{fig_illust_theory} (b); this gap can exceed the bulk band-gap as is shown in Fig. \ref{fig_illust_theory} (a); however, the SS of solution I-b does not change so much as the thickness is reduced as in Fig. \ref{fig_illust_theory} (d)-(f). Consequently, there is no apparent change in the SS even if the nature of SS is changed from I-a to I-b for systems with a finite-thickness [cf. Fig. \ref{fig_illust_theory} (a) and (b)].  These analytic results are seen to be in perfect agreements with the numerical simulation based on the model of Bi and Sb by Liu and Allen\cite{Liu1995}.

The remainder of this paper is organized as follows: in Section 2, the effective Hamiltonian for a strongly spin-orbit coupled system is introduced, and a differential equation with boundary conditions for the SS is derived; in Section 3, the results for a semi-infinite boundary condition are shown, while those for a system with finite-thickness are given in Secion 4 --- up to this point, the analytical solution have been general so they can be applied to a variety of systems with strong SOCs; in Section 5, we apply our analytical results to Bi$_{1-x}$Sb$_x$ ($0\le x \lesssim 0.1$) and compare them with numerical results; Section 6 discusses the experimentally observed SSs of Bi$_{1-x}$Sb$_x$ based on our results; and Section 7 concludes this paper.

\section{Model}
It is known that the effective Hamiltonian for a bulk system with strong SOC is generally given by the Wolff Hamiltonian\cite{Cohen1960,Wolff1964}, which is essentially to the same as the Dirac Hamiltonian but with spatial anisotropy\cite{Fuseya2015}. The Wolff Hamiltonian is obtained by applying $\kp$ theory only to the conduction and valence  bands of a system, i.e., a two-band model. With this approximation, physical quantities such as the diamagnetic orbital susceptibility of Bi can be quantitatively calculated\cite{Fukuyama1970,Fuseya2015}. However, contributions from other bands, which are neglected in the original Wolff Hamiltonian, can in some instances play crucial roles\cite{Fuseya2015b}.  The SS that originates from the SOC is one such example. In order to take into account the contributions from the other bands, in the present study we extend the Wolff Hamiltonian by using the L\"owdin partitioning up to the second order\cite{Lowdin1951,Winkler_text}. The resulting ``extended" Wolff Hamiltonian can be written as in the following form
\begin{align}
	\scr{H}=
	\begin{pmatrix}
		\D + \epsilon'(\bp) & i\bp \cdot \left[ \sum_\mu \bW (\mu) \sigma_\mu \right] \\
		-i\bp \cdot \left[ \sum_\mu \bW (\mu) \sigma_\mu \right] & -\D - \epsilon'(\bp)
	\end{pmatrix},
	\label{Bulk Hamiltonian}
\end{align}
where 
\begin{align}
	\epsilon' (\bp) & = \frac{\bp \cdot \bm{\alpha}'\cdot \bp}{2}
\end{align}
is the contribution from the other bands, while $\bm{\alpha}'$ is an inverse mass tensor, and $\sigma_\mu$ is the Pauli matrix. The contributions from the other bands are not generally symmetric between the conduction and valence bands. In this article, however, we begin with the above symmetric extended Wolff Hamiltonian, which is so general that it combines existing effective models for topological insulators and strongly spin-orbit coupled systems, such as HgTe quantum wells\cite{Bernevig2006}, Bi$_2$Se$_3$\cite{HZhang2009,CXLiu2010}, and SnTe\cite{Dimmock1964,Mitchell1966,Hsieh2012}. However, because of this generality, a more mathematically complex procedure is required. For example, in the previous models, $-i\hbar \nabla_z$ appears only in the $\sigma_z$ term, but it also appears in the $\sigma_x$ and $\sigma_y$ terms in our extended Wolff Hamiltonian.

\begin{figure}[tbp]
\begin{center}
\includegraphics[width=6cm]{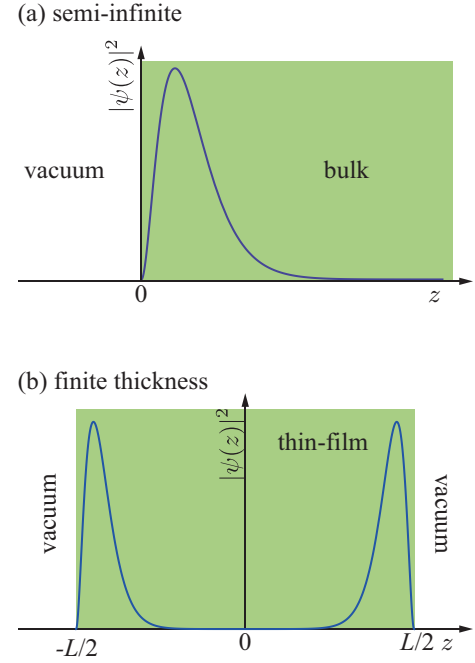}
\caption{\label{fig1} Illustration of the boundary conditions and the probability density $|\psi (z)|^2$ for (a) a semi-infinite system and (b) a system with a finite thickness. In this figure, the $z$-axis is considered to be perpendicular to the surface.}
\end{center}
\end{figure}

We analytically investigated the bound states near the surface for the Hamiltonian \eqref{Bulk Hamiltonian}. A similar approach has been used for topological insulators, such as the HgTe/CdTe quantum wells\cite{Konig2008,BZhou2008} and Bi$_2$Se$_3$ family\cite{Linder2009,HZLu2010,CXLiu2010,WYShan2010} with each effective Hamiltonian. These works only considered solutions for the topologically nontrivial SS, i.e., a SS with a massless Dirac-like dispersion. In the present study, the differential equation for the SS is solved without any assumptions being made about the topological nature of the system, i.e., both the topologically trivial and nontrivial SSs are treated equally, which allows for new SS solutions to be found.

The $z$-axis was set so as to be perpendicular to the surface, and we consider two situations: (i) a semi-infinite system in which the region $z>0$ is the bulk solid under consideration, and $z<0$ is a vacuum [Fig. \ref{fig1} (a)]; and (ii) a thin film with a finite thickness $L$, where $|z|\le L/2$ is the thin film, and $|z|>L/2$ is a vacuum  [Fig. \ref{fig1} (b)]. The Hamiltonian for these systems is commonly given by the replacement $p_z \to -i\hbar \nabla_z$. The only difference between these two situations is that of the boundary conditions used. The in-plane wave vectors $k_x $ and $k_y$ are good quantum numbers. Hereafter, $p_{x, y}$ will not be considered to be an operator, but rather an abbreviation of $p_{x, y}=\hbar k_{x, y}$. By introducing an ansatz wave function of the form $\psi \propto e^{\lambda z}$ into the extended Wolff Hamiltonian, Eq. \eqref{Bulk Hamiltonian} becomes the following form
\begin{align}
	\scr{H} =
	\begin{pmatrix}
		\D + \xi' -\frac{\alpha'_{zz}}{2}\Lambda^2 & iK_\mu \sigma_\mu \\
		-iK_\mu \sigma_\mu & -\left( \D + \xi' -\frac{\alpha'_{zz}}{2}\Lambda^2\right)
	\end{pmatrix}
	,
	\label{Surface Hamiltonian}
\end{align}
where $\lambda=\Lambda/\hbar\, (>0)$ is the inverse of the localization length, and
\begin{align}
	K_\mu 
	&= p_x W_x (\mu) + p_y W_y (\mu)  -i\Lambda W_z (\mu).
\end{align}
(The subscripts $\mu$ are assumed to be summed over pairs of identical indices.)
By separating $\epsilon' (\bp)$ into parallel $(x, y)$ and perpendicular $(z)$ components of the plane, we get the following equations
\begin{align}
	\epsilon' (\bp) &= \xi' -\frac{\alpha_{zz}'}{2}\Lambda^2,
	\\
	\xi' &=\frac{1}{2}\left( \alpha_{xx}'p_x^2 + \alpha_{yy}' p_y^2 + 2\alpha_{xy}p_x p_y\right).
	\label{xi2}
\end{align}
In these equations, tilting in the $z$-direction is neglected.

The eigenvalues of this Hamiltonian are easily obtained by considering its square, $\scr{H}^2 \psi = E^2 \psi$. By using the relation $( K_\mu \sigma_\mu )^2 = 2\D ( \xi -\alpha_{zz} \Lambda^2 /2 )$, we can obtain
\begin{align}
	E = \pm \sqrt{\left( \D + \xi' -\frac{\alpha'_{zz}}{2}\Lambda^2\right)^2 + 2\D \left( \xi -\frac{\alpha_{zz}}{2} \Lambda^2 \right) },
	\label{energy}
\end{align}
where the inverse mass and the in-plane dispersion for the conduction and valence bands are given by
\begin{align}
	\alpha_{ij}&=\frac{1}{\D}W_i (\mu) W_j (\mu),\\
	\xi &= \frac{1}{2}\left( \alpha_{xx}p_x^2 + \alpha_{yy}p_y^2 + 2\alpha_{xy}p_x p_y \right).
	\label{xi1}
\end{align}
The energy of the bulk state is given as $|E| \ge E_{\rm bulk}$, where the following equation applies
\begin{align}
	E_{\rm bulk} = \pm \sqrt{\left( \D + \xi'\right)^2 + 2\D \xi }.
\end{align}

From Eq. \eqref{energy}, we can obtain the inverse localization length as a function of $E$ in the form
\begin{align}
	\Lambda_{1, 2}^2 &= \frac{2\D}{\alpha_{zz}'^2}
	\Biggl[
	\alpha_{zz} + \alpha'_{zz} \left( 1 + \frac{\xi'}{\D} \right)
	\nonumber\\&
	\pm
	\sqrt{
	\alpha_{zz} \left( \alpha_{zz}+2\alpha'_{zz}\right)
	+\alpha_{zz}'^2 \left( \frac{E}{\D}\right)^2 
	+2\alpha'_{zz} \left( \alpha_{zz}\frac{\xi'}{\D} - \alpha'_{zz}\frac{\xi}{\D} \right)
	}
	\Biggr],
	\label{Lambda}
\end{align}
where the sign of the square root is defined to as $+$ for $\Lambda_1$ and $-$ for $\Lambda_2$. Hereafter, the unit of the energy is set to be $\D=1$.

Because the eigenvalues are doubly degenerate, the corresponding wave function can be written as a linear combination of two linearly independent eigenfunctions:
\begin{align}
	\psi_{1mn} =
	\begin{pmatrix}
		E+1+\xi' -\frac{\alpha_{zz}'}{2}\Lambda_n^2
		\\
		0
		\\
		-i(P_3 -im\Lambda_n Z_3)
		\\
		-i(P_+ -im\Lambda_n Z_+)
	\end{pmatrix}
	\label{psi1}
	,\\
	\psi_{2mn} =
	\begin{pmatrix}
		0
		\\
		E+1+\xi' -\frac{\alpha_{zz}'}{2}\Lambda_n^2
		\\
		-i \left(P_- -im\Lambda_n Z_- \right)
		\\
		i(P_3 - im\Lambda_n Z_3)
	\end{pmatrix},
	\label{psi2}
\end{align}
where
\begin{align}
	P_\mu &= p_x W_x (\mu) + p_y W_y (\mu),
	\\
	P_\pm &= P_1 \pm i P_2,
	\\
	Z_\mu &= W_z (\mu),
	\\
	Z_\pm &= Z_1 \pm i Z_2.
\end{align}
By considering the degrees of freedom of $\Lambda_{1,2}$ and $\pm$ signs in $e^{\pm \Lambda_n z/\hbar}$, the general solution for the wave function is given by a linear combination of eight eigenfunctions as\cite{WYShan2010}
\begin{align}
	\Psi(z)=\sum_{l =1,2}\sum_{m=\pm}\sum_{n=1,2}C_{lmn}\psi_{lmn}e^{m\lambda_n z}.
	\label{psi8}
\end{align}

\section{Semi-infinite system}
In this section, we focus on the solutions for the semi-infinite system. The boundary conditions for the semi-infinite system are given by the following equation:
\begin{align}
	\Psi (z=0) =\Psi (z=+\infty) = 0. 
	\label{boundary1}
\end{align}
The latter condition requires $m=-$, so that the wave function is practically given as a linear combination of four eigenfunctions
\begin{align}
	\Psi (z)&= \left( C_{1-1}\psi_{1-1} + C_{2-1}\psi_{2-1}\right) e^{-\lambda_1 z}
		\nonumber\\
		&
		+\left( C_{1-2}\psi_{1-2} + C_{2-2}\psi_{2-2} \right) e^{-\lambda_2 z}.
\end{align}
The condition $\Psi (z=0)=0$ yields the four simultaneous equations for $C_{l-n}$, 
\begin{align}
	C_{1-1}\psi_{1-1} + C_{2-1}\psi_{2-1}
	+C_{1-2}\psi_{1-2} + C_{2-2}\psi_{2-2}  =0.
	\label{4eqs}
\end{align}
This leads to the eigenfunctions in the form
\begin{align}
	\Psi(z) =  \left( C_{1-1}\psi_{1-1} + C_{2-1}\psi_{2-1}\right) \left( e^{-\lambda_2 z} - e^{-\lambda_1 z} \right).
\end{align}
The depth dependence of the probability density $|\psi (z)|^2$ of the wave function is illustrated in Fig. \ref{fig1} (a). The wave function is localized very close to the surface and decays exponentially in the bulk region. The peak position depends on the difference between $\lambda_1$ and $\lambda_2$.

The secular equation of Eq. \eqref{4eqs} is the following:
\begin{align}
	\begin{vmatrix}
		\left( E+ 1+\xi'-\frac{\alpha_{zz}'}{2}\Lambda_1^2\right) I
		&
		\left( E+ 1+\xi'-\frac{\alpha_{zz}'}{2}\Lambda_2^2\right) I
		\\
		-i (P_\mu +i\Lambda_1 Z_\mu) \sigma_\mu
		&
		-i (P_\mu +i\Lambda_2 Z_\mu) \sigma_\mu
	\end{vmatrix}
	=0,
	\label{secular1}
\end{align}
where $I$ is a $2\times 2$ unit matrix. From this secular equation, we can obtain the following equation with respect to $E$:
\begin{align}
	\left(\Lambda_1 -\Lambda_2\right)^2
	\left[
	\frac{\alpha'^2_{zz}}{4} \xi\left(\Lambda_1+\Lambda_2\right)^2
	-\frac{\alpha_{zz}}{2}
	\left( E+1+\xi' +\frac{\alpha_{zz}'}{2}\Lambda_1 \Lambda_2 \right)^2
	\right] 
	\nonumber\\
	=0
	\label{eq_semi}
\end{align}
By substituting Eq. \eqref{Lambda} into Eq. \eqref{eq_semi}, we obtain an equation to solve in the following form
\begin{align}
\left(E+1+\xi'-\beta \xi \right)
\left\{
1+\xi'+{\rm sgn}(\beta) \sqrt{(1+\xi')^2+2\xi - E^2}
\right\}=0.
\label{eq1}
\end{align}
(The factor $(\Lambda_1-\Lambda_2)$ is discarded since the solution with $\Lambda_1 =\Lambda_2$ gives $\Psi (z)=0$.)
At this point, we introduced an important parameter
\begin{align}
	\beta = \frac{\alpha'_{zz}}{\alpha_{zz}},
\end{align}
which characterizes the relative curvature of the other band contributions to that of the conduction and valence bands in the direction of the $z$-axis. Its sign is closely related to the topological property of the model\cite{SQShen_book}, but there is no exact relationship between them. 
The solutions of Eq. \eqref{eq1} are as follows
\begin{align}
	E_{\rm s1\pm}&=\pm \sqrt{2\xi}, 	\tag{Solution I}
	\\
	E_{\rm s2\pm}&=\pm\left( 1+\xi'-\beta \xi\right). \tag{Solution II}
\end{align}
(The solution of $E_{\rm s2 +}$ is obtained by using two linearly independent eigenfunctions similar to Eqs. \eqref{psi1} and \eqref{psi2} but include $E-1-\xi' +\frac{\alpha'_{zz}}{2}\Lambda_n^2$.)

\subsection{Solution I}
\begin{figure}[tbp]
\begin{center}
\includegraphics[width=6cm]{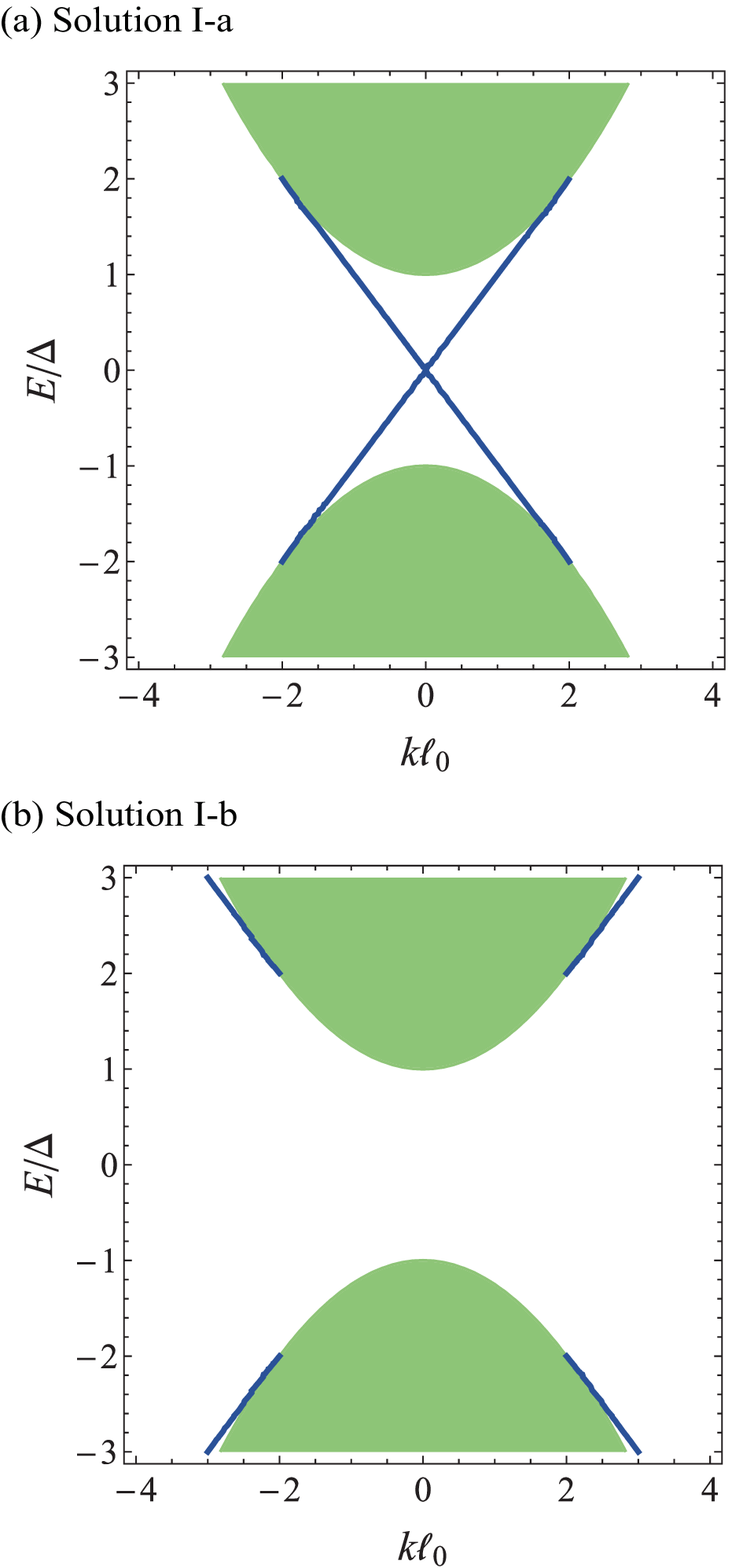}
\end{center}
\caption{\label{fig_sol1a} (a) Solution I-a ($\beta=-0.1$, $\alpha_\parallel=1.0/m$, $\alpha'_\parallel=-0.5/m$) and (b) solution I-b ($\beta=0.1$, $\alpha_\parallel=1.0/m$, $\alpha'_\parallel=-0.5/m$) of the extended Wolff Hamiltonian. The shaded regions indicate the two-dimensional bulk bands.
}
\end{figure}
There are two types of solution I. Their existence conditions depend on the sign of $\beta$ as
\begin{align}
	\mbox{(I-a):}\,\,\,\, \beta <0 \mbox{ and } 1+\xi' >0, 
	\\
	\mbox{(I-b):}\,\,\,\,\beta>0 \mbox{ and } 1+\xi'<0. 
\end{align}
Although both solution I-a and I-b have the same dispersion of $\pm \sqrt{2\xi}$, which is linear with $k$, the regions in which they appear are different, as shown in Fig. \ref{fig_sol1a}. In the following figures, the in-plane dispersion is set to be $\xi = \alpha_{\parallel}\hbar^2 k^2/2$ and $\xi' = \alpha'_{\parallel}\hbar^2 k^2/2$, which does not lose any generalities. The wave number and the localization length are scaled in terms of $\ell_0 = \hbar/\sqrt{m\D}$ (in the case of Bi with $2\D=13.6$meV,\cite{Liu1995} $\ell_0$ is estimated to be $\ell_0 =33.3$\AA, which is about eight layers. The thickness of one Bi bilayer is 3.94\AA.\cite{Hofmann2006})

Solution I-a appears around $k=0$, and the linear dispersions cross at $k=0$, i.e., they cross as the massless Dirac dispersion. Note that the origin of $k$ is not necessarily at the center of the Brillouin zone; it can be any extremum of the band as the conventional $\kp$ theory.  

On the other hand, solution I-b can only appear away from $k=0$ since $1+\xi' <0$ is never satisfied near $k=0$. The existence of solution I-b has not been mentioned in previous works\cite{Konig2008,BZhou2008,Linder2009,HZLu2010,CXLiu2010,WYShan2010,SQShen2012,SQShen_book}; it has been recognized that no SS appears for $\beta>0$. The existence of solution I-b is shown for the first time in the present work.


The important quantity of a SS is its localization length. We can calculate the inverse of the localization length $\Lambda$ by substituting the solution into Eq. \eqref{Lambda}. The inverse localization length of solution I is obtained as
\begin{align}
	\Lambda_{1, 2}^2=\frac{2}{\alpha_{zz}\beta^2}
	\left[
	1+\beta (1+\xi') \pm \sqrt{1+2\beta(1+\xi')}
	\right].
	\label{localization1}
\end{align}
The dispersion of solution I enters into the bulk state at $\xi'=-1$. At this $k$-point, $\Lambda_2$ becomes zero, and the SS naturally continues to the bulk state. It is interesting to note that the energy dispersion depends only on $\xi$, while the localization length depends only on $\xi'$. 

\begin{figure}[tbp]
\begin{center}
\includegraphics[width=7cm]{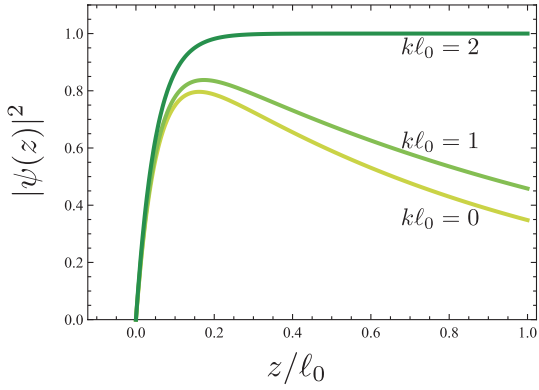}
\caption{Probability of the wave function $|\psi (z) |^2$ for solution I-a as a function of $z$ with $\beta=-0.1$, $\alpha_\parallel=1.0/m$, $\alpha'_\parallel=-0.5/m$. It should be noted that the SS enters into the bulk band at $k\ell_0 = 2.0$.}
\label{fig7}
\end{center}
\end{figure}

The probability of the wave function $\psi (z)\equiv e^{-\Lambda_2 z/\hbar}-e^{-\Lambda_1 z/\hbar}$ of solution I-a is shown in Fig. \ref{fig7} for different values of $k\ell_0$ in the case where $\beta=-0.1$, $\alpha_\parallel=1.0/m$, $\alpha'_\parallel=-0.5/m$. The wave function is mostly bound at $z / \ell_0 \simeq$0.1--0.2, i.e., the first and second layers in the case of Bi. In this case, the SS enters into the bulk state at $k\ell_0=2$, where the localization length is infinitely long (Fig. \ref{fig7}), i.e., it naturally continues to the bulk wave function. The properties of $|\psi (z)|^2$ for solution I-b are essentially the same as those for solution I-a.

\subsection{Solution II}
\begin{figure}[tbp]
\begin{center}
\includegraphics[width=8.0cm]{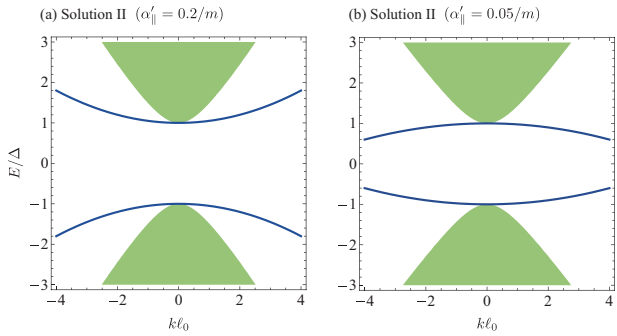}
\end{center}
\caption{\label{fig2} Solution II for (a) $\alpha'_\parallel=0.2/m$, and (b) $\alpha'_\parallel=0.05/m$ with $\beta=0.1$, $\alpha_\parallel=1.0/m$. The shaded regions indicate the two-dimensional bulk band dispersions.
}
\end{figure}

The energy dispersion of solution II is tangential to $E_{\rm bulk}$ at the band edge $E=\pm 1$ and $k_{x, y}=0$. When $\xi'-\beta \xi =0$, the surface band is perfectly flat. When $\xi' -\beta \xi \neq 0$, the surface band has a finite mass that depends on the value of $\xi'-\beta \xi$; it can be positive [Fig. \ref{fig2} (a)] or negative [Fig. \ref{fig2} (b)]. 

\begin{figure}[tbp]
\begin{center}
\includegraphics[width=7cm]{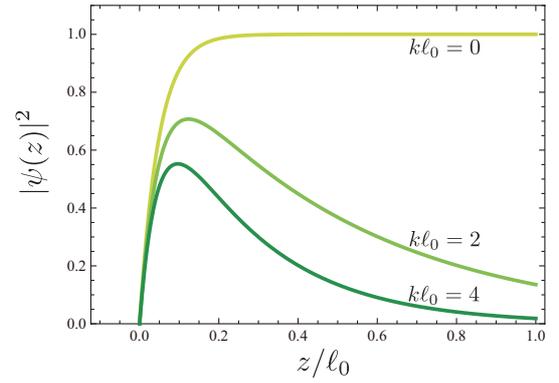}
\caption{\label{fig3}Probability of the wave function, $|\psi (z, k) |^2$, of solution II as a function of $z$ with $\beta=0.1$, $\alpha_\parallel=1.0/m$, $\alpha'_\parallel=0.1/m$. $\alpha_{zz}$ is set to be $1.0/m$.}
\end{center}
\end{figure}

For solution II, the inverse localization length can be obtained as
\begin{align}
	\Lambda_{1} &=
		\frac{2}{\sqrt{\alpha_{zz}}|\beta|}\sqrt{1+\beta \left( 1+\xi' \right) -\frac{\beta^2}{2}\xi},
		\\
	\Lambda_{2} &=\sqrt{\frac{2\xi}{\alpha_{zz}}}.
	\label{localization_length}
\end{align}
In the limit of $k\to0$ ($\xi, \xi'\to0$), where $E_{\rm s2}$ enters into the bulk state, $\Lambda_2$ becomes zero, and the localization length becomes infinitely long. The probability of the wave function, $|\psi (z)|^2$, of solution II is shown in Fig. \ref{fig3} as a function of the depth $z$. The properties of $|\psi (z)|^2$ are essentially common to the properties obtained by another set of parameters.

Note that the solution II with a nearly flat dispersion resembles the SS (or the edge state) with flat band in the graphene ribbons\cite{Nakada1996,Bernevig_book} or in the ``line node semimetals" by Burkov--Hook--Balents.\cite{Burkov2011} They share a common feature that the SS enters into the bulk state at the band edge. On the other hand, there is no specific condition for the appearance of solution II, while there is a certain condition for that in line node semimetals.

\section{Finite-thickness system}
In this section, we consider a finite-thickness system. The boundary conditions for a system with thickness $L$ is given by the following equation:
\begin{align}
	\Psi (z=+L/2)=\Psi(z=-L/2)=0.
	\label{boundary_finite}
\end{align}
The wave function is given as a linear combination of eight eigenfunctions \eqref{psi8}. For example, the terms including $C_{1+1}$ and $C_{1-1}$ are explicitly written as
\begin{align}
	&\begin{pmatrix}
		\left(E+1+\xi' -\frac{\alpha_{zz}'}{2}\Lambda_1^2 \right)
		\left\{ D_1 \cosh (\lambda_1 z) + D_2 \sinh(\lambda_1 z) \right\}
		\\
		0
		\\
		-\Lambda_1 Z_3
		\left\{ D_1 \sinh (\lambda_1 z) + D_2 \cosh(\lambda_1 z) \right\}
		\\
		-\Lambda_1 Z_+ 
		\left\{ D_1 \sinh (\lambda_1 z) + D_2 \cosh(\lambda_1 z) \right\}
	\end{pmatrix}
	\nonumber\\
	&+\begin{pmatrix}
	0
	\\
	0
	\\
	-iP_3 \left\{ D_1 \cosh (\lambda_1 z) + D_2 \sinh(\lambda_1 z) \right\}
	\\
	-iP_+ \left\{ D_1 \cosh (\lambda_1 z) + D_2 \sinh(\lambda_1 z) \right\}
	\end{pmatrix},
\end{align}
where $D_1 = C_{1+1}+C_{1-1}$ and  $D_2 = C_{1+1}-C_{1-1}$. Similarly, the terms including $C_{2+1}$ and $C_{2-1}$ are given as
\begin{align}
	&\begin{pmatrix}
		0
		\\
		\left( E+1+\xi' -\frac{\alpha_{zz}'}{2}\Lambda_1^2 \right)
		\left\{ D_3 \cosh (\lambda_1 z) + D_4 \sinh(\lambda_1 z) \right\}
		\\
		-\Lambda_1 Z_- 
		\left\{ D_3 \sinh (\lambda_1 z) + D_4 \cosh(\lambda_1 z) \right\}
		\\
		\Lambda_1 Z_3 
		\left\{ D_3 \sinh (\lambda_1 z) + D_4 \cosh(\lambda_1 z) \right\}
	\end{pmatrix}
	\nonumber\\
	&+
	\begin{pmatrix}
		0
		\\
		0
		\\
		-iP_- \left\{ D_3 \cosh (\lambda_1 z) + D_4 \sinh(\lambda_1 z) \right\}
		\\
		iP_3 \left\{ D_3 \cosh (\lambda_1 z) + D_4 \sinh(\lambda_1 z) \right\}
	\end{pmatrix}.
\end{align}
The rest of the terms can be obtained by simply changing $n=1\to2$ and their coefficients $D_{1\sim4} \to D_{5\sim 8}$ correspondingly in the above form. The boundary condition yields the eight simultaneous equations for $D_{1\sim 8}$ and its secular equation $(8\times 8)$ is given by the following equations:
\begin{align}
	|L|&=
	\begin{vmatrix}
		L_1 & L_2 \\
		L_3 & L_4
	\end{vmatrix}
	=0,
\end{align}
\begin{align}
	L_1 &= 
	\begin{pmatrix}
		E_1 C_1 & 0 & E_1 S_1 & 0
		\\
		0 & E_1 C_1 & 0 & E_1 S_1
		\\
		-\Lambda_1 Z_3 S_1 & -\Lambda_1 Z_- S_1 & -\Lambda_1 Z_3 C_1 & -\Lambda_1 Z_- C_1
		\\
		-\Lambda_1 Z_+ S_1 & \Lambda_1 Z_3 S_1 & -\Lambda_1 Z_+ C_1 & \Lambda_1 Z_3 C_1
	\end{pmatrix}
	\nonumber\\
	&+
	\begin{pmatrix}
	0 & 0 & 0 & 0 
	\\
	0 & 0 & 0 & 0
	\\
	-iP_3 C_1 & -i P_- C_1 & -iP_3 S_1 & -i P_- S_1
	\\
	-iP_+ C_1 & iP_3 C_1 & -iP_+ S_1 & iP_3 S_1
	\end{pmatrix}
	\nonumber\\
	&=
	\begin{pmatrix}
		E_1 C_1 I & E_1 S_1 I
		\\
		-\Lambda_1 S_1 Z_\mu \sigma_\mu & -\Lambda_1 C_1 Z_\mu \sigma_\mu
	\end{pmatrix}
	+
	\begin{pmatrix}
	0 & 0
	\\
	-i C_1 P_\mu \sigma_\mu & -i S_1 P_\mu \sigma_\mu
	\end{pmatrix},
	\\
	L_2 &= L_1 (n = 1\to 2),
	\\
	L_3 &= L_1 (S_1 \to -S_1),
	\\
	L_4 & = L_2 (S_1 \to -S_1),
\end{align}
where $E_n = E+1+\xi' -\frac{\alpha'_{zz}}{2}\Lambda_n^2$, $C_n = \cosh(\lambda_n L/2)$, and $S_n = \sinh(\lambda_n L/2)$. This leads to the $z$-dependence of the eigenfunction as
\begin{align}
	\Psi(z) &=
	F_1 \left\{\frac{\cosh (\lambda_1 z)}{\cosh (\lambda_1 L/2)}
	-\frac{\cosh (\lambda_2 z)}{\cosh (\lambda_2 L/2)} \right\}
	\nonumber\\
	&+F_2 \left\{\frac{\sinh (\lambda_1 z)}{\sinh (\lambda_1 L/2)}
	-\frac{\sinh (\lambda_2 z)}{\sinh (\lambda_2 L/2)} \right\},
\end{align}
where $F_1$ and $F_2$ are coefficients of four components. This functional form satisfies the boundary conditions given in Eq. \eqref{boundary_finite} (an example of this functional form is illustrated in Fig. \ref{fig1} (b)).

The determinant of the matrix $L$ is given by
\begin{align}
	|L| &=
	16\Bigl[
	2\xi (E_1 -E_2)^2 C_1 C_2 S_1 S_2
	\nonumber\\
	&
	-\alpha_{zz}(E_1 \Lambda_2 C_1 S_2 - E_2 \Lambda_1 C_2 S_1)
	(E_1 \Lambda_2 C_2 S_1 - E_2 \Lambda_1 C_1 S_2)
	\Bigr]^2.
\end{align}
Here, we used the following relations:
\begin{align}
	\sum_\mu P_\mu^2 &=2\D \xi,
	\\
	\sum_\mu Z_\mu^2 &= \D \alpha_{zz},
\end{align}
\begin{align}
	&4(P_1 P_2 Z_1 Z_2 + P_2 P_3 Z_2 Z_3 + P_3 P_1 Z_3 Z_1)
	\nonumber\\
	&+P_1^2 (Z_1^2 -Z_2^2 -Z_3^2)
	+P_2^2 (-Z_1^2 + Z_2^2 -Z_3^2)
	+P_3^2 (-Z_1^2 -Z_2^2 + Z_3^2)
	\nonumber\\
	&=\D^2 \Bigl[
	(2\alpha_{zx}^2-\alpha_{zz}\alpha_{xx})p_x^2
	+(2\alpha_{yz}^2-\alpha_{yy}\alpha_{zz})p_y^2
	\nonumber\\&
	+2(2\alpha_{zx}\alpha_{yz}-\alpha_{xy}\alpha_{zz})p_x p_y
	\Bigr].
\end{align}
One possible difficulty in utilizing the Wolff Hamiltonian is that the parameters $\bm{W}(\mu)$, $P_\mu$, and $Z_\mu$ are complex in general, which may cause a phase dependent result in numerical calculations. By using the above relations, however, we were able to succeed in writing the equation only in terms of the mass tensors, which are not affected by the phase of the wave functions. This is practically important for the numerical calculations.

Following this, we can obtain the equation with respect to $E$:
\begin{align}
	&2\xi (E_1 -E_2)^2 T_1 T_2
	\nonumber\\
	&
	-\alpha_{zz}(E_1 \Lambda_2 T_2 - E_2 \Lambda_1 T_1)
	(E_1 \Lambda_2 T_1 - E_2 \Lambda_1 T_2)
	=0,
	\label{eq_finite}
\end{align}
where $T_n = \tanh (\lambda_n L/2)$. In the limit of $L\to + \infty$, where $T_1, T_2 \to 1$, Eq. \eqref{eq_finite} becomes equivalent to Eq. \eqref{eq_semi}; the solutions of Eq. \eqref{eq_finite} should therefore naturally continues to those of the semi-infinite system.

\subsection{Solution I}
\begin{figure}[tbp]
\begin{center}
\includegraphics[width=6cm]{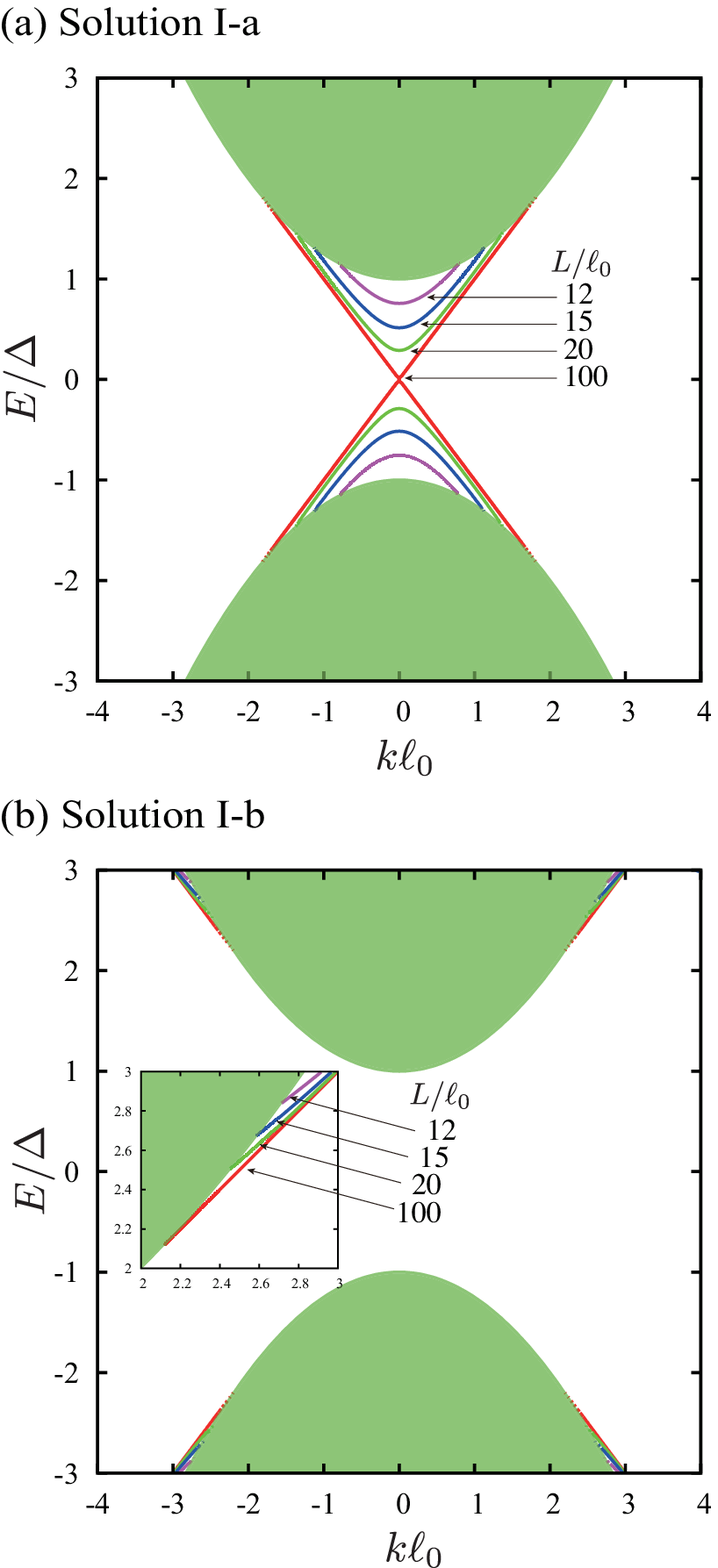}
\end{center}
\caption{\label{fig_sol1a_finite} 
Solution I for the system that has a finite thickness, $L/\ell_0 = 12, 15, 20$, and $100$, $\alpha_{zz}=100/m$, $\alpha_\parallel=1.0/m$, and $\alpha'_\parallel=-0.5/m$. 
(a) Solution I-a with $\alpha'_{zz}=-1/m$ ($\beta=-0.01$). (b) Solution I-b with $\alpha'_{zz}=1/m$ ($\beta=0.01$). The inset in (b) is an enlarged image of the point where the SS enters into the bulk band.
}
\end{figure}
\begin{figure}[tbp]
{\small (a)}
\begin{center}
\includegraphics[width=7cm]{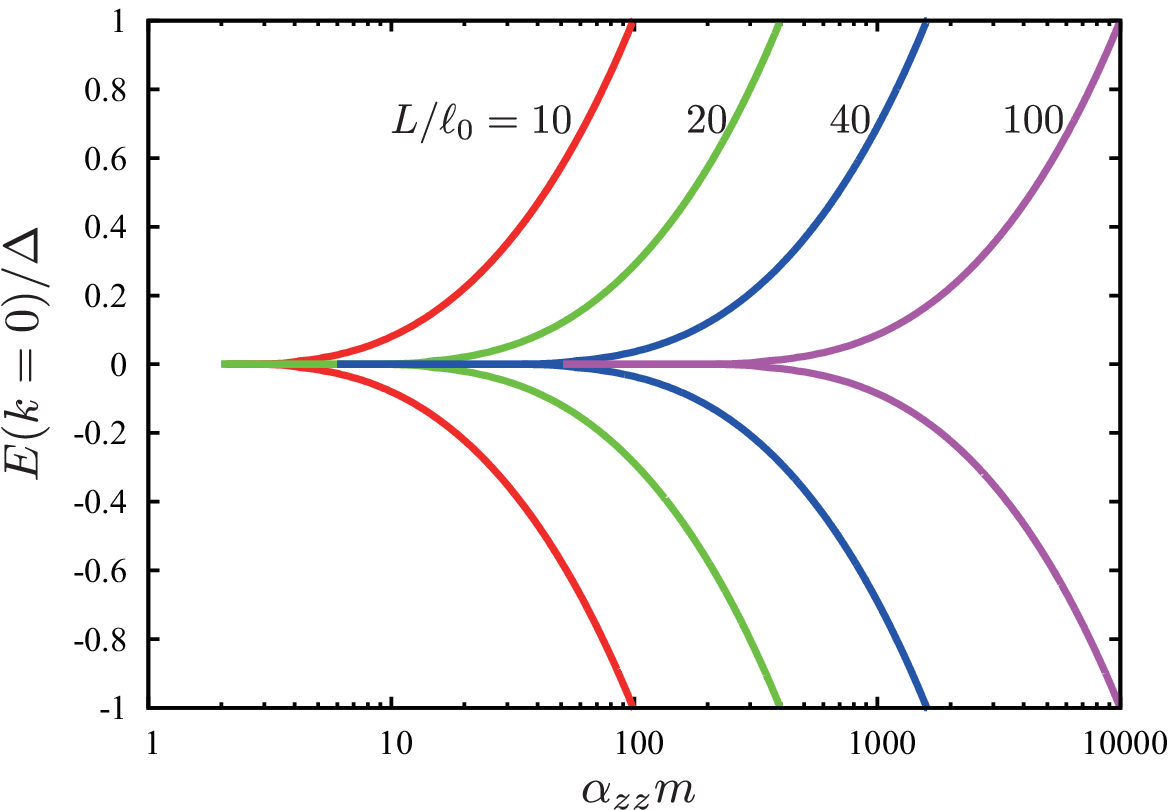}
\end{center}
{\small (b)}
\begin{center}
\includegraphics[width=7cm]{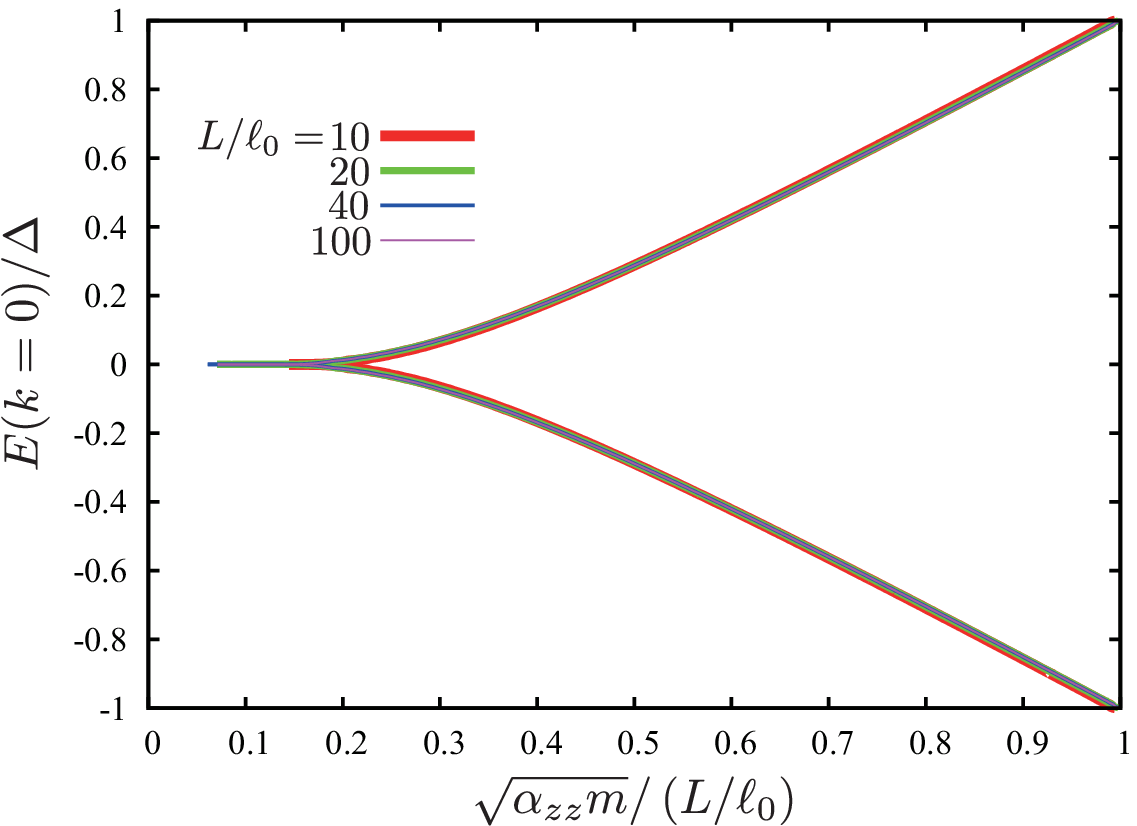}
\end{center}
\caption{\label{fig_gap} 
Energy of solution I-a at $k=0$ as a function of (a) $\alpha_{zz}m$, and (b) $\sqrt{\alpha_{zz}}/(L/\ell_0)$ for $L/\ell_0=10, 20, 40$, and $100$ and $\alpha'_{zz}=-1.0/m$. The energy between two lines with the same $L$ corresponds to the energy gap of solution I-a.}
\end{figure}

Figure \ref{fig_sol1a_finite} shows the dispersion of solution I-a and I-b for different thicknesses $L/\ell_0 = 12, 15, 20$, and $100$. We can see that the linearly crossing SSs of I-a open the gap as the thickness decreases; this is due to the interference between the opposite sides of the surfaces, which is also found in other systems\cite{BZhou2008,Linder2009,HZLu2010,SQShen_book,Ozawa2014}.
Figure \ref{fig_gap} shows the $\alpha_{zz}$ dependence of the energy of I-a at $k=0$, indicating the size of the gap, $E_g$. Interestingly, it is found that the size of the gap is scaled with respect to $\sqrt{\alpha_{zz}}/L$ as shown in Fig. \ref{fig_gap} (b), i.e., it is independent from $\alpha'_{zz}$. As $\alpha_{zz}$ increases (the effective mass decreases) and $L$ decreases, $E_g$ increases. The gap of the SS becomes as large as the size of the bulk band gap, $E_g = 2\D$, when $\sqrt{\alpha_{zz}m}=L/\ell_0$.

Solution I-b also exhibits a thickness dependence, though its effect is not so remarkable. The energy of the point at which the SS enters the bulk band increases as the system becomes thinner, as shown in the inset of Fig. \ref{fig_sol1a_finite} (b).

The localization length is almost the same as that obtained for the semi-infinite system, since the $E$-dependence of $\Lambda_n$, Eq. \eqref{Lambda},  is common between the semi-infinite and finite-thickness systems, though there is a slight modulation due to the opening of the gap.

\subsection{Solution II}
We mentioned that the equation with respect to $E$ for the finite-thickness system, Eq. \eqref{eq_finite}, naturally becomes that of the semi-infinite one in the limit of $L\to +\infty$. If we take the opposite limit, $L\to 0$, Eq. \eqref{eq_finite} becomes the following equaion:
\begin{align}
	\frac{\alpha_{zz}\alpha'_{zz}L^2}{8\hbar^2}
	\Lambda_1 \Lambda_2 \left( \Lambda_1^2 -\Lambda_2^2 \right)^2
	\left( E+1 + \xi'-\beta \xi \right)=0.
\end{align}
To be more precise, solution II is also a solution in the zero-thickness limit. This indicates that there is no quantum size effect for solution II, which is similar to the edge state of graphene ribbons with a flat band\cite{Nakada1996}. However, such a solution does not appear in the numerical calculation in the next section, and does not seem to appear in experiments. We will therefore only concentrate on solution I-a and I-b hereafter.

\section{Comparison with the numerical simulations}
\begin{figure}[tbp]
{\small (a)}
\begin{center}
\includegraphics[width=5cm]{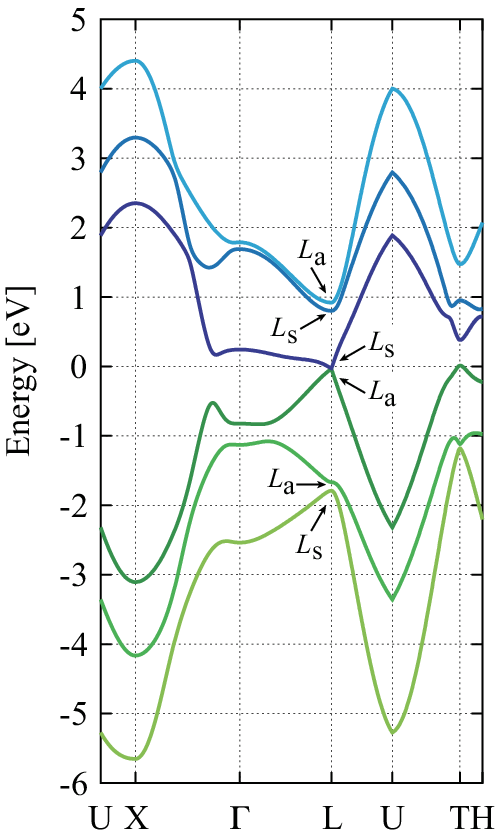}
\end{center}
{\small (b)}
\begin{center}
\includegraphics[width=7cm]{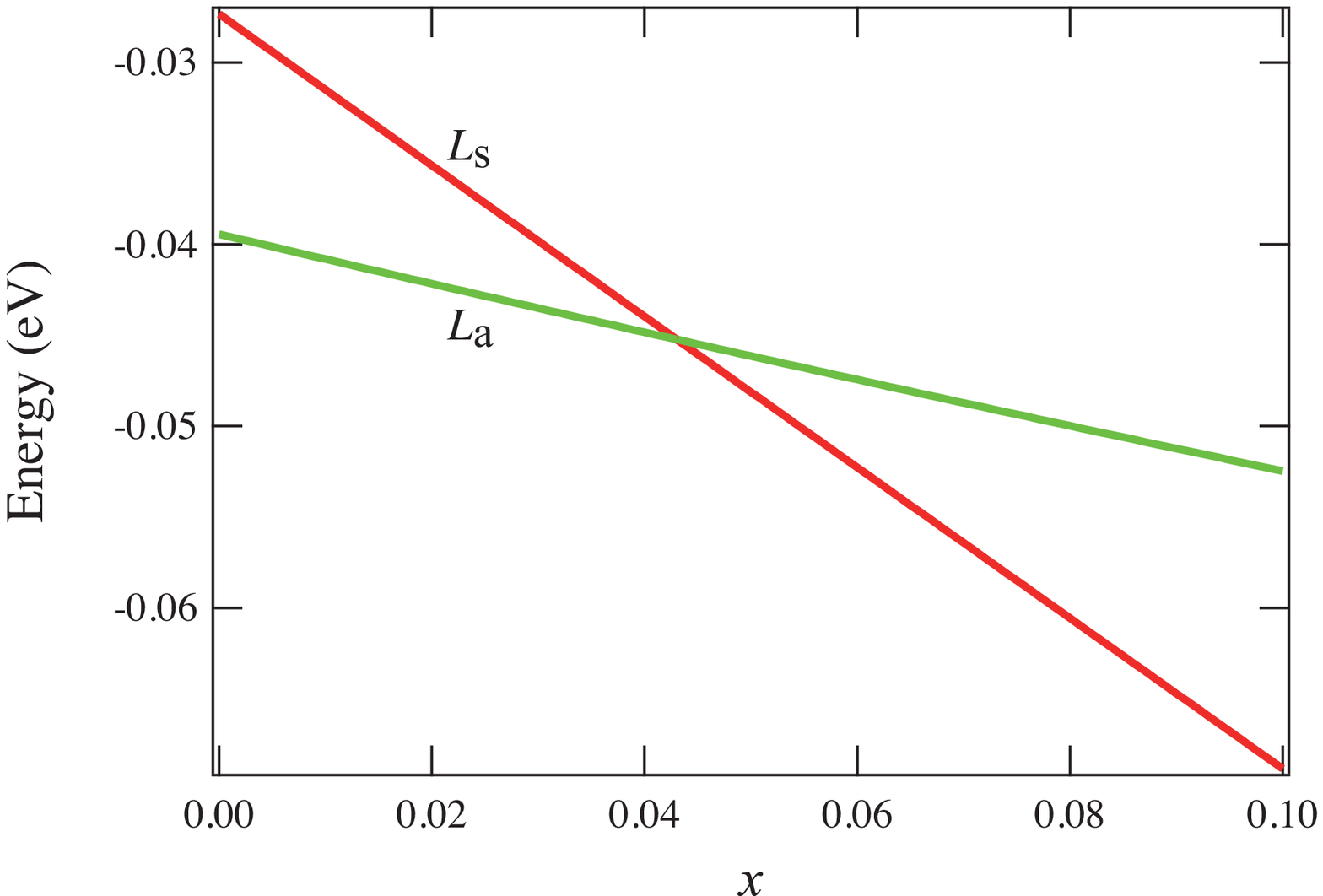}
\end{center}
\caption{\label{fig_band} (a) Band structure of bulk Bi and the symmetries at the $L$ point as calculated by the tight-binding model proposed by Liu and Allen\cite{Liu1995}. (b) Evolution of the conduction and valence bands at the $L$ point with respect to the Sb concentration, $x$, based on the VCA. The theoretical value of $x$ is scaled up by a factor of 2 so that the comparison with the results from experiments can be done in a more straightforward manner\cite{Fuseya2015b}.
}
\end{figure}
\begin{figure}[tbp]
{\small
(a) pure Bi
}
\begin{center}
\includegraphics[width=8cm]{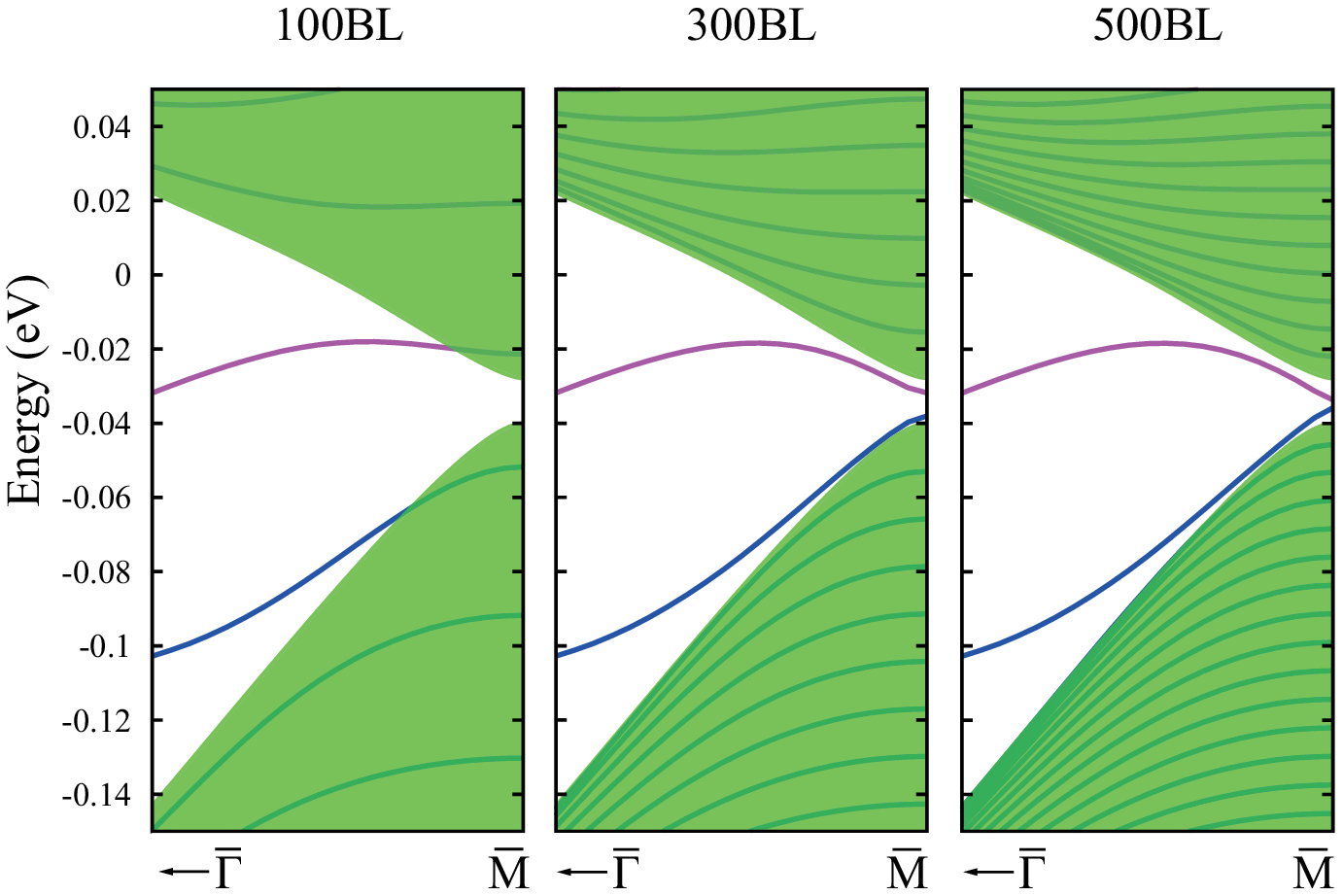}
\end{center}
{\small
(b) Bi$_{0.9}$Sb$_{0.1}$
}
\begin{center}
\includegraphics[width=8cm]{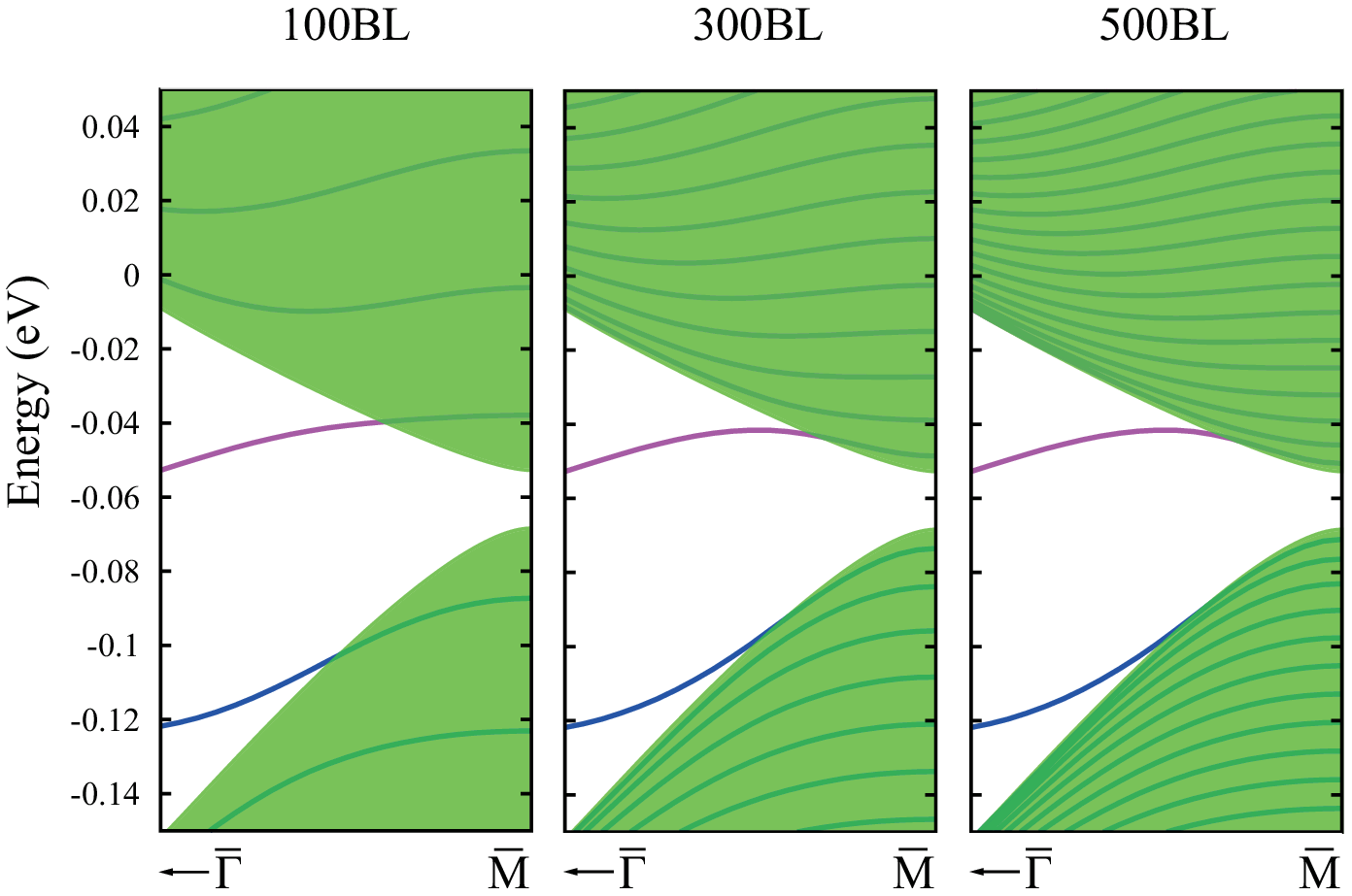}
\end{center}
\caption{\label{fig_finiteTB} Band structures of (a) pure Bi and (b) Bi$_{0.9}$Sb$_{0.1}$ for 100, 300, and 500 bilayer (BL) systems, as calculated using the Liu--Allen model\cite{Liu1995,Fuseya2015b}. The shaded region corresponds to the bulk band. The origin of the energy is fixed to being that of the Fermi energy of pure Bi for both (a) and (b). Note that pure Bi is topologically trivial and Bi$_{0.9}$Sb$_{0.1}$ is topologically nontrivial for the Liu--Allen model according to the classification of Fu and Kane\cite{Fu2007}.}
\end{figure}

In this section, we will compare our analytical solutions with the numerical solutions using the tight-binding model of Liu and Allen\cite{Liu1995} for Bi$_{1-x}$Sb$_x$ ($0\le x \le 0.1$). Figure \ref{fig_band} shows the obtained band structure and its symmetry at the $L$ point. The order of the band, from its bottom to its top, is $\{ L_{\rm s}, L_{\rm a}, L_{\rm a} , L_{\rm s}, L_{\rm s}, L_{\rm a}\}$. This order is classified to be topologically ``trivial", according to Fu and Kane.\cite{Fu2007} For the substitution of Sb, we adopt a simple virtual crystal approximation (VCA), in which the tight-binding parameters are obtained by a linear extrapolation between pure Bi and Sb. The band evolution obtained in this manner agrees well with the results of previous experimental investigations\cite{Jain1959,Lerner1968,Tichovolsky1969,Brandt1970,Buot1971,Brandt1972,Oelgart1976,Vecchi1976,Lenoir1996}. The band inversion at the $L$ point occurs at $x=0.043$ according to the present calculation --- this can be seen in Fig. \ref{fig_band} (b) --- and its topology changes from trivial to nontrivial at this point\cite{Fu2007}. It should be noted that the Sb content here $x$ is scaled up by a factor of 2 in order for comparisons with the experimental results to be more straightforward. The details of the calculations are described in a previous study\cite{Fuseya2015b}.

\begin{figure}[tbp]
\begin{center}
\includegraphics[width=7cm]{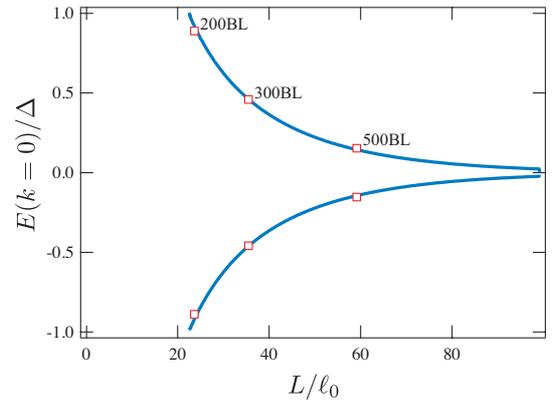}
\end{center}
\caption{\label{fig_Ldependence} Energy of SSs at $k=0$ as a function of the thickness $L$. The solid lines are the analytical results with $\alpha_{zz}=498.7/m$ and $\alpha'_{zz}=-3.554/m$. The symbols are the numerical results obtained using the Liu--Allen model. The thickness of a 1 BL system is set to being 3.94 $\AA$ and $\ell_0=33.3$ \AA.\cite{Hofmann2006}}
\end{figure}

Figure \ref{fig_finiteTB} shows the energy dispersion of the finite bilayer (BL) system around the $\bar{M}$ point based on the Liu--Allen model. For both Bi and Bi$_{0.9}$Sb$_{0.1}$, there are two SSs around the $\bar{M}$ point; however, their thickness dependences are different. For pure Bi, one of the SSs enters into the bulk conduction band, while the other enters into the valence band of a 100 BL system. In 300 and 500 BL systems, however, the SSs do not enter the bulk band and tend to cross with one another in the direct gap of the bulk band. This is a characteristic feature of solution I-a. The thickness dependence of the energy at $k=0$ for a finite BL system agrees perfectly with the analytical solution we obtained using Eq. \eqref{eq_finite} at $k=0$ (which is shown in Fig. \ref{fig_Ldependence}). For the analyticalal solution, the input parameters are only $\alpha_{zz}=498.7/m$ and $\alpha'_{zz}=-3.554/m$, which are obtained from the ``bulk" calculation of the Liu--Allen model. The perfect agreement between them indicates the correctness and validity of the analytical solution. This is the first time that the nature of the SS in the numerical calculation is clarified with the help of analytical solusions. It is rather surprising that even the 500 BL system exhibits a gap due to the interference between both side surfaces, i.e., 500 BL ($\simeq 200$ nm thickness) cannot be regarded as the bulk system\cite{Ohtsubo2016}. This remarkable quantum size effect is due to the small bulk band-gap and the small effective mass of Bi.

\begin{figure}[tbp]
\begin{center}
\includegraphics[width=7cm]{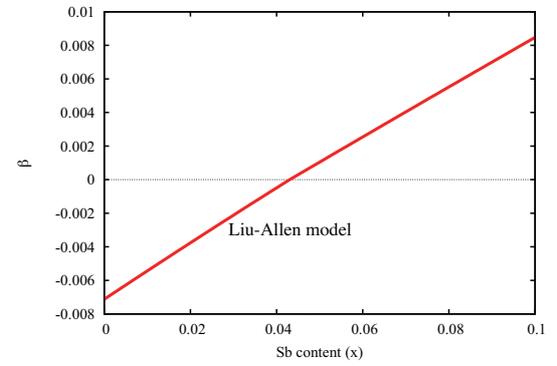}
\\\,\vspace{5mm}
\end{center}
\caption{\label{fig_doping} Sb content dependence of $\beta=\alpha'_{zz}/\alpha_{zz}$ obtained by using the Liu--Allen model with a VCA. 
}
\end{figure}

For Bi$_{0.9}$Sb$_{0.1}$, however, the SSs enter into the bulk bands even in the 500 BL system. They are considered to be solution I-b; if so, the sigh of $\beta$ of the ``bulk" Bi$_{1-x}$Sb$_x$ should change from negative to positive due to the band inversion.
The Sb content dependence of $\beta$, which is based on the Liu--Allen model, is plotted as a solid line in Fig. \ref{fig_doping}. $\beta $ can be seen in this figure to change its sign from negative to positive, which corresponds to the qualitative change in SS from solution I-a to I-b. This is a clear evidence for the qualitative transition of the SS due to the band inversion. Therefore, as far as based on the Liu-Allen model, the sign change in $\beta$ accompanies the band inversion. 

The largest contribution to $\beta$ for the conduction band at the $L$ point in pure Bi is that of the $L_{\rm a}$ band located at 0.9 eV shown in Fig. \ref{fig_band} (a). It may be rather surprising that the contributions from such a high energy band more than 1 eV far from the Fermi energy drastically affect the appearance SS. However, this can happen in the strongly spin-orbit coupled system, in which the large SOC ($\sim 1.8$ eV in Bi) produces a significant interband effect\cite{Fuseya2015b}.

The most important consequence is that the appearance of the SS in the finite-thickness system is essentially unchanged by the band inversion even if the true nature of the SS is changed. For example, when we compare the 100 BL system results of pure Bi and Bi$_{0.9}$Sb$_{0.1}$ in Fig. \ref{fig_finiteTB}, we cannot distinguish between them by only looking at their dispersions; rather, the differences only appear in very thick BL or semi-infinite systems. According to the thickness dependence exhibited in Fig. \ref{fig_Ldependence}, the difference can be seen in the system with more than 200 BL ($\simeq 80$ nm) at least.

\section{Discussions}
\subsection{Comparison with the experimental results}

First, we briefly summarize the experimental results on Bi$_{1-x}$Sb$_x$ ($0\le x \lesssim 0.1$). The bulk band inversion in Bi$_{1-x}$Sb$_x$ has been examined repeatedly more than five decades through various types of measurements, such as magneto-optics\cite{Tichovolsky1969,Oelgart1976,Vecchi1976}, transport and quantum oscilations\cite{Jain1959,Lerner1968,Brandt1970,Brandt1972}, and magnetic susceptibility\cite{Buot1971}. Nowadays, there is no doubt that the conduction and valence bands at the $L$ point are inverted at around $x\simeq 0.04$.

The previous experiments on (111) surface of Bi$_{1-x}$Sb$_x$\cite{Jezequel1986,Hengsberger2000,Ast2001,Ast2003,Koroteev2004,Hofmann2006,Hirahara2006,Hirahara2007,Ohtsubo2013,Hirahara2015,Benia2015,Ito2016} are summarized as follows. There are two SSs between the $\bar{\Gamma}$ and $\bar{M}$ points. (The $\bar{\Gamma}$ point corresponds to the $\Gamma$-$T$ line in the bulk, and $\bar{M}$ corresponds to the $L$-$X$ line.) Near the $\bar{\Gamma}$ point, both of the SSs enter the bulk valence band, and their properties are unchanged for $0\le x \lesssim 0.1$. Near the $\bar{M}$ point, one SS enters into the bulk conduction band, while the other SS enters into the valence band. Even though the bulk conduction and valence bands at the $L$ point are inverted, the properties of SSs at around the $\bar{M}$ point are essentially unchanged.
The finite-thickness\cite{Hirahara2006,Hirahara2007,Nakamura2011,Benia2015,Ito2016} and the semi-infinite\cite{Hengsberger2000,Ast2001,Ast2003,Ohtsubo2013} systems exhibit these properties in common. 
The essence of these experiments is the fact that the SSs seem to remain qualitatively unchanged through the band inversion, at which the topology of the system is changed. 
The results of present analytical studies supported by numerical calculations are in complete agreement with these experimental indications.

As has been discussed in the previous sections, when the thickness is finite, a gap opens in solution I-a. In the case of Bi$_{1-x}$Sb$_x$, because the bulk band-gap and the bulk band-mass are very small, this quantum size effect appears to be drastic enough that the gap of the SS exceeds the gap of the bulk-band. In reality, the gap of the SS in a 200 BL system is comparable to the bulk band gap, according to the estimation we made. (Note that the most recent measurements were carried out up to 202 BL for pure Bi\cite{Ito2016} and up to about 300 BL for $x>0.07$\cite{Benia2015}.) Therefore, for the system with finite thickness ($\lesssim 200$), solution I-a for $x\lesssim 0.04$ cannot be distinguished from solution I-b for $x\gtrsim 0.04$. Therefore, the properties of SSs are apparently unchanged by the band inversion. 

The difference between solution I-a and I-b should appear in the semi-infinite system. When the band inversion is accompanied by a sign change in $\beta$, there should be a qualitative transition from solution I-a to I-b in case of the bulk surface. If $\beta <0$ in pure Bi, as is estimated by the  Liu--Allen model, then linearly crossing SSs should be observed before the band inversion ($0\le x \lesssim 0.04$). However, such a gapless SS has not yet been observed, so there still remains an inconsistency between our numerical simulations based on the Liu--Allen model and the experiments on the bulk surface.

This discrepancy between our theoretical results and the experimental results for the bulk SS can be resolved if the resolution of the surface ARPES is improved. In pure Bi, the bulk band gap is 11--15 meV, and it is much smaller in Bi$_{1-x}$Sb$_x$. On the other hand, the current experimental resolution of surface ARPES was 25 meV in the study conducted by Ast and H\"ochst\cite{Ast2001,Ast2003}, and it was 7 meV in the study conducted by Ohtsubo {\it et al.}\cite{Ohtsubo2013}, which are not high enough for Bi$_{1-x}$Sb$_x$.

It is worth noting the recent progress made for ARPES measurements of the SS of the iron-based superconductor FeTe$_{1-x}$Se$_{x}$\cite{PZhang2017}. Although the first-principles calculations predict a bulk band-gap of $\sim 10$ meV and the linearly crossing SSs, such a SS has not been identified by ARPES measurements that had energy resolutions of $\sim 5$ meV and $\sim 10$ meV\cite{XShi2017}. However, by using high energy and momentum resolution ARPES with an energy resolution of $\sim 70$ $\mu$eV, linearly crossing SSs and the bulk band-gap were clearly detected\cite{PZhang2017}. Similar progress is therefore expected for Bi$_{1-x}$Sb$_x$.

\subsection{Some remarks on the topology of the system}

To determine the topology of the system, we need to take into account the entire Brillouin zone; especially the parity eigenvalue at eight time-reversal invariant momenta.
On the other hand, since our theory is based on $\kp$ theory, which is rigorous for the local region in the Brillouin zone, we cannot give a statement on the topology of the system. Nevertheless, we can make some remarks on the topology of the system based on our results as follows.

In some cases, the topology of the system has been determined only by counting the intersections of the SSs and the Fermi energy. However, argument of such kind can lead to an incorrect solution for the system with finite thickness, since the opening the gap of SS due to the interference between opposite sides of surfaces can change the number of intersections, even though the topology of the system is unchanged.
In connection with this, the conclusion that the pure Bi must be topologically nontrivial since one of the two SSs enters into the bulk conduction band and the other enters into the bulk valence band near the $\bar{M}$ can be incorrect for the system with finite thickness. There is a possibility that even topologically trivial system exhibits such SSs entering conduction and valence bands separately as is shown in Fig. \ref{fig_finiteTB} (a).

\section{Conclusion}
We have succeeded in obtaining the exact solutions for SSs of both semi-infinite and finite-thickness systems described by the extended Wolff Hamiltonian, which is common to strongly spin-orbit coupled systems. There are three types of solutions for the semi-infinite system; solution I-a and I-b have a dispersion linear in $k$, and solution II has a nearly flat dispersion.

In the semi-infinite system, solution I-a appears when $\beta<0$ and $1+\xi'>0$. The dispersion of solution I-a linearly cross with each other at $k=0$, where $k$ is measured from an extremum of the bulk band. On the other hand, solution I-b appears when $\beta>0$ and $1+\xi'<0$. One of the SSs of I-b linearly enters into the bulk conduction band, and the other enters into the bulk valence band away from $k=0$. Solution I-a corresponds to the gapless Dirac-like SS that has previously been discussed for topological insulators. Solution I-b is a finding that has been made for the first time.
The key parameter that classifies the nature of a SS (solution I-a or I-b) is the sign of $\beta$. 

In a finite-thickness system, a gap opens in solution I-a due to the interference between opposite sides of the surface. The size of the gap is scaled by $\sqrt{\alpha_{zz}}/L$, and it becomes comparable to the bulk band gap for $\sqrt{\alpha_{zz}m}=L/\ell_0$. As a result, this quantum size effect is more significant for smaller effective masses and bulk band gaps.

We have also calculated the SS of Bi$_{1-x}$Sb$_x$ for $0\le x \le 0.1$ based on the tight-biding model of Liu and Allen. We have obtained a perfect agreement between the analytical and numerical results: the thickness dependence of the gap of solution I-a, and the qualitative change in the SS from solution I-a to I-b through the band inversion at $x\simeq 0.04$. Due to this agreement, the nature of the numerically obtained SS has been identified for the first time based on the analytic solutions. For a system with a thickness below about 200 BL ($\sim$80 nm), the gap in solution I-a exceeds the bulk band gap. As a result, solution I-a becomes qualitatively the same as solution I-b. This can give a possible answer to the experimental mystery on the (111) surface of Bi$_{1-x}$Sb$_x$ for the finite-thickness system. However, there still remains discrepancy between our results and experiments for the semi-infinite system. For this discrepancy, high-resolution ARPES measurements will give a crucial progress.

Although we focused on the problem on Bi$_{1-x}$Sb$_x$ in this paper, the present analytical solutions are not restricted to this specific case. They are applicable to various systems with strong SOC, including IV-VI semiconductors (PbTe, SbSe, SnTe), and topological insulators.

\vspace{1cm}
\begin{acknowledgment}
	We greatly appreciate fruitful discussions with T. Hirahara. Y. F. would like to thank N. Sasaki, J. Nakamura, and K. Hattori for helpful discussions. 
	Y. F. is supported by JSPS KAKENHI Grants No. 16K05437, No. 15H02108, and No. 25870231.
\end{acknowledgment}


\bibliographystyle{jpsj}
\bibliography{Bismuth}

\end{document}